\documentclass[aps,prl,preprint,groupedaddress]{revtex4-1}
\usepackage{amsmath,amsfonts,amsthm,amssymb}  
\usepackage{url} 
\usepackage[inline]{trackchanges} 
\usepackage{graphicx} 
\date{\today}
\usepackage{hyperref}
\usepackage{cleveref}
\usepackage{array,booktabs}
\newcolumntype{L}{@{}>{\kern\tabcolsep}l<{\kern\tabcolsep}}
\usepackage{colortbl}
\usepackage{xcolor}
\begin{document}


\title{Cross-Diffusion Waves as a Mesoscopic Uncertainty Relationship for Multi-Physics Instabilities} 

\author{K. Regenauer-Lieb}
\affiliation{School of Minerals and Energy Resources Engineering, UNSW, Sydney, NSW 2052 Australia }
\author{ M. Hu}
\affiliation{Department of Civil Engineering, The University of Hong Kong, Hong Kong }
\affiliation{School of Minerals and Energy Resources Engineering, UNSW, Sydney, NSW 2052 Australia }
\author{C. Schrank}
\affiliation{Science and Engineering Faculty, Queensland University of Technology,  Brisbane, QLD, 4001, Australia }


%

\begin{abstract}
We propose a generic uncertainty relationship for cross-diffusion (quasi-soliton) waves triggered by local instabilities through Thermo-Hydro-Mechano-Chemical (THMC) coupling and cross-scale feedbacks. Cross-diffusion waves nucleate when the overall stress field is incompatible with accelerations from local feedbacks of generalized THMC  thermodynamic forces with generalized thermodynamic fluxes of another kind. Cross-diffusion terms in the 4 x 4 THMC diffusion matrix are shown to lead to multiple diffusional $P$- and $S$-wave solutions of the coupled THMC equations.  Uncertainties in the location of local material instabilities are captured by wave scale correlation of probability amplitudes. Cross-diffusional waves have unusual dispersion patterns and, although they assume a solitary state, do not behave like solitons but have a quasi-elastic particle-like state. Their characteristic wavenumber and constant speed defines mesoscopic internal material time-space relations entirely defined by the coefficients of the coupled THMC reaction-cross-diffusion equations. These coefficients are identified here as material parameters underpinning the criterion for nucleation and speed of diffusional waves.  Interpreting patterns in nature as features of standing or propagating diffusional waves offers a simple mathematical framework for analysis of multi-physics instabilities and evaluation of their uncertainties similar to their quantum-mechanical analogues. 
\end{abstract}

\maketitle

\section{Introduction}
In a paper presented at the Nobel Conference XXVI, St. Peter, in Minnesota on the 2-3 Oct 1990 Ilya Prigogine postulated that we are at the beginning of an era of  'New Physics' \citep{Prigogine}. He proposed that systems that are pushed far from equilibrium react by forming new space/time dissipative structures that lead to internal material time scales. He goes on to suggest that incorporating this new time scale at the interface of solving the problems of the physics of the very small to the very large defines a new frontier in science. 
 
This contribution extends on a first multi-scale and multi-physics formulation that specifically considers the internal time scale associated with the development of resonant, coupled oscillators through a cross-diffusion approach \citep{Regenauer2019}. In this contribution we consider, without loss of generality, as example multiphysics Thermo-, Hydro-, Mechanical- and Chemical (THMC) coupling. The approach is inspired by the quantum-mechanical uncertainty relationship here proposed to apply for larger-scale dissipative structures. 

\section{Uncertainty through Cross-Diffusion in the Quantum Realm}
In this upscaled wave mechanics approach of the quantum limit we postulate that the uncertainty is related to the nucleation and subsequent evolution of cross-diffusion waves. We will show that this cross-diffusional uncertainty is caused by the discrete nature of energy eigenstates (dissipative structures) that appear as the equivalent of the reduced Planck's quantum $\hslash$ at larger scale. We will illustrate the proposed analogy by interpreting $\frac{\hslash^2}{2m}$ as a cross-diffusion coefficient at quantum level. 

The concept of a cross-diffusion wave  at quantum level for a particle in a quantum state travelling through a space without potential can be illustrated by using a decomposition of the diffusive probability amplitude of a complex wave function. Such a particle is characterized by the simple potential-free equation 

\begin{equation}
\label{eq:Schroedinger1}
 i \hslash \frac{\partial{\psi(x, t)}}{\partial{t}} = -\frac{\hslash^2 \nabla^2}{2m} \psi(x, t) ,
\end{equation}

where $\psi$ denotes the complex position-space wave function. Assuming $\psi(x,t)=u(x,t)+iv(x,t)$, we obtain a cross-diffusion type relationship where the probability amplitude of the $u$-wave depends on the cross-diffusional coupling to the $v$-wave. Vice versa, the probability amplitude of the $v$-wave depends on the cross-diffusion of $u$:

\begin{subequations}
\label{eq:Schroedinger2}
\begin{align}
\hslash \frac{\partial{u(x, t)}}{\partial{t}} = -\frac{\hslash^2 \nabla^2}{2m} v(x, t) \\[0.2in]
\hslash \frac{\partial{v(x, t)}}{\partial{t}} = \frac{\hslash^2 \nabla^2}{2m} u(x, t) .
\end{align}
\end{subequations}

This Cauchy-Riemann decomposition of complex variables allows an interesting  interpretation of the proposed bound quantum state of larger-scale discrete dissipative structures. If indeed the complex position-space wave function Eq. (\ref{eq:Schroedinger1}) applies to larger-scale quantized energy states it would suggest that cross-diffusion coefficients yield probability amplitudes whose squared modulus provides the probability density distribution of local dissipative structures (such as local material failure) as an uncertainty relationship in terms of coupling between the real and imaginary part of a wave function.

This meso-scale uncertainty relationship is proposed here as a multi-scale formalism. We identify Prigogine's internal material time-scale as the time scale defined by propagating cross-diffusion waves and start with molecular-scale processes to encounter discrete quantized energy states at different scales, which are defined by the chemical reactions and mass exchange processes under thermodynamic driving forces, triggering THMC thermodynamic fluxes. The product of thermodynamic forces and fluxes defines irreversible entropy production in terms of Prigogine's dissipative structures \citep{Kondepudi}. These discrete structures define the various quantized energy states at larger than quantum scale and form the basis for the uncertainty relationship discussed here. In a thermodynamic sense this  meso-scale uncertainty relationship underpins the upper and lower bound principles of entropy production \citep{Regenauer2010}. 

\section{Quantum-state dissipative structures}
Before discussing cross-scale coupling it is useful to briefly review insights into the formation of quantum-state dissipative structures. The concept was introduced first in chemical and biological systems where morphogenic patterns \citep{Turing} were identified as solutions to the underlying reaction-diffusion equations. These discrete patterns were later on named Turing patterns. An excellent update of the formulation for chemical systems can be found in \citep{Vanag}.
 
\begin{figure}
\centering
\includegraphics[width=.6\textwidth]{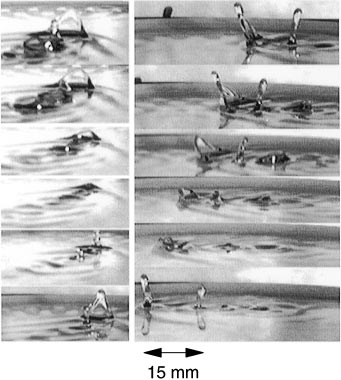}
\caption{ Water molecules exhibit a discrete quantum-like solitary state when forced by a mechanical shaker at a critical condition (here 41 Hz). Periodic finger-like solitary states travel from right to left  at a constant velocity. Each snapshot shows 20 ms intervals. Unlike classical solitons their appearance is particle-like. They can pass through each other with a slight loss of amplitude, or “collide” to create a new state whose direction of propagation is at an angle to that of the original states (image from ref \citep{DissipativeAccelerationWave}).}
\label{fig:solitarystate}
\end{figure}

The approach has been generalized here for the solid mechanics application where experimental evidence is rare \citep{Hu2019b} due to the recent nature of the proposition. For the fluid mechanics of surface-driven fluids experimentalists have, however,  shown (see Fig. \ref{fig:solitarystate}) that as the system becomes highly dissipative, it can undergo a sharp transition from a continuum state to one of highly localized, propagating, particle-like state \citep{DissipativeAccelerationWave}. We propose here that these observations are universal for THMC reaction-diffusion systems that are driven far from equilibrium. The approach allows an interpretation of these observations in terms of propagating particle-like states which emerge as stationary Turing patterns for long-time-scale standing wave solution of the cross-diffusion formulation.

These discrete patterns lead to conditions of violation of smoothness on surfaces in the body, where one or multiple internal variables from the lower scale suffer jump discontinuities owing to the discrete nature of upscaled quantum-like states. This is the physical reason for the formation of acceleration fronts, where the cross-diffusive length-scales $\sqrt{L_{ij}t}$ ($L_{ij}$ are the cross diffusion coefficients, see Eq. \ref{eq:crossdiffusionmatrix}) are quantized  \citep{DissipativeAccelerationWave} and linked to the convective velocity of the step function on cross-diffusion waves. The discrete nature of particle-like states as the physical reason for emergence of cross-diffusion waves provides the formal analogy to the above discussed uncertainty relationship at quantum level.

A similar observation of energy quantization twelve orders of magnitude larger than Planck's quant has been found in space plasmas  \citep{Plasmon}. In such plasmas clusters of particles aggregate into Debye spheres  at a characteristic correlation length-scale given by the Debye length. Evidence for this large scale energy quantization can be found in the wave spectrum of space plasmas, where a distinct plasmon oscillation frequency is found in analogy to the oscillation frequency of a photon \citep{Plasmon}.

Here we postulate that distinct particle-like energy eigenstates exist at multiple scales. The quantized energy eigenstate at macro-scale forms through long-range interactions of multiphysics quasi-soliton waves \citep{Tsyganov2014} that propagate themselves at meso-scale, as illustrated in Fig.(\ref{fig:solitarystate}), in a quantized particle-like solitary state. We identify these quasi-solitons \citep{Biktashev2016} as cross-diffusion waves and focus on formulating an approach for the meso-scale physics of these cross-diffusion waves to arrive at a macro-scale continuum mechanics extension of the Navier-Stokes equation valid for solids and fluids. Before discussing the thermodynamics of THMC cross-scale coupling and the emergence of cross-diffusion waves, we first present a generalized continuum constitutive approach and separate shear and volumetric deformation. 

\section{Wave Equations}

\subsection{Constitutive Assumptions}
\label{ch:continuum}
The fundamental equation of motion is:

\begin{equation}
\label{eq:A1}
\nabla \cdot \boldsymbol{\sigma}+\rho \,\mathbf{f}=\rho \,\mathbf{a}    ,
\end{equation}

where $\boldsymbol{\sigma}={{\sigma }_{ij}}$ is the Cauchy stress tensor, $\rho $ the density, $\mathbf{f}$ is a body force (e.g. gravity) and  $\mathbf{a}$ the acceleration. This equation does not stipulate a constitutive law but with constitutive assumptions it becomes the master equation for the theory of elastic waves, fluid mechanics and continuum mechanics.  Material properties can be expressed, by using the Helmholtz decomposition, in terms of shear ($S$-wave) and compressional ($P$-wave) wave velocities which is a convenient description for the purpose of this paper.

For an isotropic elastic medium, for instance, accelerations in Eq. (\ref{eq:A1}) are only allowing elastic displacements described by $\mathbf{u}$. In this case the material can be characterized by just two velocities: the elastic $P$-wave velocity $v_p$ and the elastic $S$-wave velocity $v_s$, and we obtain from Eq. (\ref{eq:A1}) the elastic-wave equation:

\begin{equation}
\frac{\partial^2  \mathbf{u}}{\partial t^2} = \underbrace{v_p^{2} \nabla(\nabla \cdot \mathbf{u})}_{P \text { wave }}-\underbrace{v_s^{2} \nabla \times(\nabla \times \mathbf{u})}_{S \text { wave }}  .
\label{eq:elasticwave}
\end{equation}
 
Similarly, by allowing the material to deform in a viscous manner, where accelerations are characterised by velocities $\mathbf{v}$, the Helmholtz decomposition identifies a scalar $P$-wave and a vectorial $S$-wave potential field, where the material constants are the dynamic shear $\eta$ and bulk $\zeta$ viscosities to obtain the Navier-Stokes equation:

\begin{equation}
\label{eq:NavierStokes}
 \rho \,\left( \frac{\partial \mathbf{v}}{\partial t}+\mathbf{v}\cdot \nabla \mathbf{v} \right)=-\nabla p+2{{\nabla }^{2}} (\eta \boldsymbol{\dot{\epsilon}}')+\nabla(\zeta(\nabla \cdot \mathbf{v}))+\rho \,\mathbf{f} 
\end{equation}

with 
$$\boldsymbol{\dot \epsilon_0} = \frac{1}{3}(\nabla \cdot \mathbf{v}) \mathbf{I}$$
where  $\mathbf{I}$ is the identity matrix and 
$$\boldsymbol{\dot \epsilon}'=\frac{1}{2}\left(\nabla \mathbf{v}+(\nabla \mathbf{v})^{T}\right)-\boldsymbol{\dot \epsilon_0}
$$ being the deviatoric viscous strain rate.

Here, we use the thermodynamic definition of the pressure as $p=-\frac{\partial U}{\partial V}$ where $U$  is the internal energy and  $V$ the volume.  In the present context pressure is therefore defined as $p=\frac{1}{3} tr({\boldsymbol{\sigma}})$ and is negative for compression. 

For the elasto-visco-plastic case we have the equivalent fourth-order elasto-visco-plastic stiffness tensor $\mathbf{C}$ characterizing material stiffness and the corresponding elasto-visco-plastic bulk viscosity $\zeta$ to give

\begin{equation}
\label{eq:viscoplastic}
 \rho \,\left( \frac{\partial \mathbf{v}}{\partial t}+\mathbf{v}\cdot \nabla \mathbf{v} \right)=-\nabla p+2\nabla (\boldsymbol{C}{{\boldsymbol{{\dot{\epsilon}}'}}})+3 \nabla(\zeta {\dot \epsilon_0})+\rho \,\mathbf{f} ,
 \end{equation}

in this case $\boldsymbol{\dot \epsilon}'$ denotes the deviatoric elasto-visco-plastic strain rate and ${\dot \epsilon_0}$ the equivalent volumetric strain-rate. For simplicity, we assume here a Maxwellian rheology implying a separation of elastic and visco-plastic wave time-scales in the context of an additive strain-rate decomposition of Eq. (\ref{eq:elasticwave}) and (\ref{eq:viscoplastic}). In the following we only discuss the slow visco-plastic wave phenomenon.

%
%

\subsection{Acceleration Waves in the Creeping Flow Regime}
\label{ch:appendix2}	
\begin{figure}
\centering
\includegraphics[width=.6\textwidth]{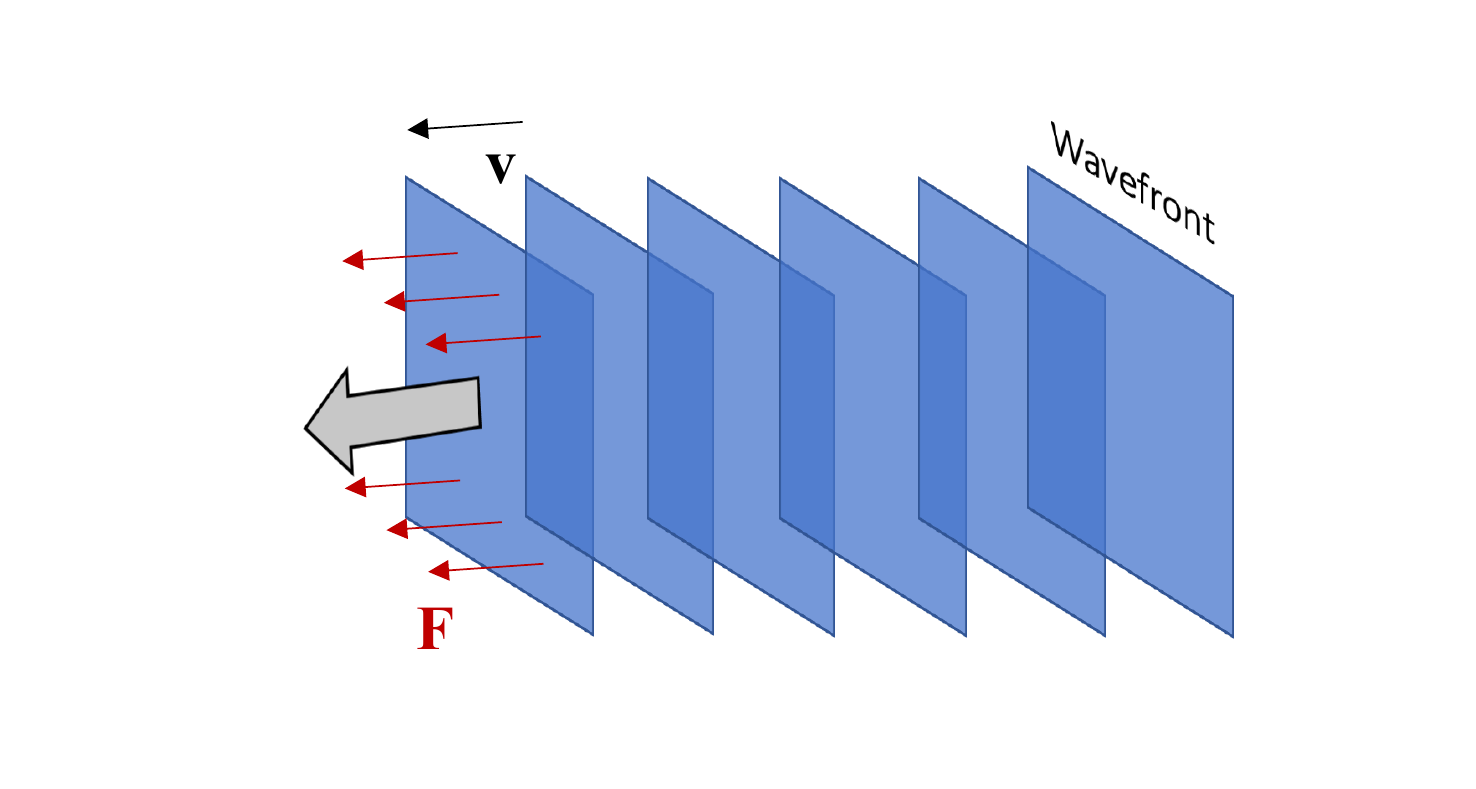}
\caption{ Acceleration waves can originate at a body surface when the existing internal stress gradient is dynamically incompatible with accelerations imposed on particles of the surface. A propagating plane-wave front is shown here for reference but a plane-wave is not a necessary restriction. Across these surfaces  particle accelerations and spatial gradients of velocity  are momentarily discontinuous while the velocity itself is continuous. }
\label{fig:acceleration}
\end{figure}

While fluid- and solid-wave phenomena obviously occur when the above equation includes inertial forces, so-called '\textit{acceleration waves}' \citep{Coleman}, caused by local surfaces of acceleration (see Fig. \ref{fig:acceleration}), can also occur in the  creeping flow limit  when no acceleration due to a (gravity) potential is present and $\rho \,\mathbf{f}=0$. This formulation implies that only the kinetic Hamiltonian operator is active, thus providing an analogy to the wave-function for a quantum state particle travelling through a space without  potential formulated in equation (\ref{eq:Schroedinger1}). These acceleration waves are defined as geometric surfaces  (here assumed to be plane-waves) that move relatively to the material \citep{Hill_acceleration}.  

Acceleration waves can be described in two ways. One can use two coordinate systems, one for the reference state and one for the current state. A more elegant way is to consider convective coordinate systems by formulating the constitutive law in terms of stress rate. For this we consider the space derivative normal to the moving wavefront (see Figure \ref{fig:acceleration}) indicated by $\frac{\partial}{\partial s}$. Waves are travelling with respect to a background Lagrangian moving  material reference frame. The relativistic reference frame therefore becomes a segment of the moving wavefront $\frac{\partial}{\partial s}$ with the time progressing normal to the wavefront. 

Considering the traction (load per unit area) in the direction normal to the wavefront as $\mathbf{F}$  and choosing the magnitude of velocity of the moving wavefront as $c$ , the jump condition indicated by the square Iverson brackets can be advected along $c$. This leads to Hadamard's jump condition where the true traction rate along the advected coordinates is:

\begin{equation}
\label{eq:A11}
\left[ \mathbf{\dot F}\right]  = - c \left[\frac{\partial {\mathbf{F}}}{\partial s}\right] .
\end{equation}

Hadamard's jump condition applies to all internal variables and the acceleration across the wavefront is constrained by 

\begin{equation}
\label{eq:A12}
 \left[\rho \mathbf{\dot{v}}\right] = \left[\frac{\partial {\mathbf{F}}}{\partial s}\right].
\end{equation}

Combining Eq. (\ref{eq:A11}) and (\ref{eq:A12}) we obtain

\begin{equation}
\label{eq:A13}
\textbf{[}\mathbf{\dot{F}}\textbf{]} +   c \textbf{[} \rho\mathbf{\dot{v}}\textbf{]} = 0 .
\end{equation} 

Substituting  $\mathbf{\dot{v}}=c \frac{\partial \mathbf{ v}}{\partial s}$ into Eq. (\ref{eq:A13}) we obtain

\begin{equation}
\label{eq:HillEinstein}
\textbf{[}\mathbf{\dot{F}}\textbf{]} +  c^2  \left[\rho \frac{\partial \mathbf{{v}}}{\partial s} \right] = 0 .
\end{equation}

Similar to Einstein's relativistic correlation of energy to matter ($E=m c^2$ ) Hill's formulation of acceleration waves in Eq. (\ref{eq:HillEinstein}) expresses the energetics of the acceleration waves by the square of the material velocity times the mass of the characteristic segment defined by $\frac{\partial \mathbf{{v}}}{\partial s}$. This provides a simple formulation where the energetics of the material is solely described by Eq. (\ref{eq:A1}) and the mesoscale mass exchange rates on acceleration waves in  Eq. (\ref{eq:HillEinstein}). The material velocity $c$, being the velocity of acceleration waves, becomes a material constant for propagation of acceleration waves.

Acceleration waves form the basis of localization criteria in plasticity theory. The criterion for instability is derived from the equivalent theory in elastodynamics where for an elasto-plastic body the acoustic tensor  $\boldsymbol \Gamma$ is defined by

\begin{equation}
\label{eq:acoustic tensor}
\boldsymbol \Gamma = \mathbf{n} \cdot \mathbf{C} \cdot \mathbf{n}
\end{equation} 

In elastodynamics the eigenvalues of $\boldsymbol \Gamma$ divided by the mass density represent the squares of the elastic wave propagation speed in the direction of the unit normal vector $\mathbf{n}$. In elasto-plasticity the equivalent dynamic stability criterion is defined by Eq. (\ref{eq:HillEinstein}) which in terms of acoustic tensor implies that 
 
\begin{equation}
\label{eq:elastoplastic stability}
\boldsymbol {\Gamma} \frac{\partial \mathbf{{v}}}{\partial s}  = \rho c^2 \frac{\partial \mathbf{{v}}}{\partial s} 
\end{equation} 

Dynamic system stability can be evaluated through assessing the eigenvalues of the acoustic tensor thus determining the speed of the acceleration waves which must be real. Mathematically,  Eq. (\ref{eq:elasticwave} - \ref{eq:viscoplastic}) can be represented by the addition of two functions, a scalar field and the curl of a vector field. The former without curl or rotation identifies compressional $P$-waves and the latter features zero divergence and corresponds to isochoric sinistral and dextral shear $S$-waves. 

These dissipative waves are interpreted as stationary (standing) waves when the determinant of the acoustic tensor is zero, 

\begin{equation}
\label{eq:acoustic tensor}
det(\boldsymbol \Gamma) = 0
\end{equation} 

which is the standard condition for localisation in plasticity theory \citep{Vardoulakis_book}. Accordingly, formation of localized shear-bands out of homogeneous plastic flow is assumed when the velocity of the wavefront vanishes.  Hill's criterion for localisation \citep{Hill_acceleration} was mainly concerned with shear acceleration waves in an ideal linear time-independent  elasto-plastic material where two families of characteristics (dextral and sinistral slip lines)  feature a jump in strainrate at the wavefront accompanied by one in stress rate (but not in stress). This in turn implies a related jump in stress gradient.  Later work extended the theory to formulate accelerations waves as a the basis of modern criteria for localization in plastic media \citep{Rudnicki75,Rice76}. In those theories the possibility of volumetric acceleration waves was, however, neglected and volumetric deformation was parameterized by an empirical dilatancy angle. 

Another shortcoming of the localization criterion for the application to THMC instabilities is that it is not directly applicable to the rate-dependent elasto-visco-plastic case. The inclusion of rate effects implies a positive wave speed different from zero \citep{Perzyna1996}. To date, no generally accepted localization criterion for localisation phenomenon for  Eq. (\ref{eq:viscoplastic}) exists although stationary localization phenomena for rate-sensitive materials are clearly observed in the laboratory and nature. The method of choice to date is to use all field equations and perform a numerical stability analysis. A discussion on an extension to the above discussed criterion has been presented recently \citep{Pisano} and a review of energy based criteria that successfully model the adiabatic limit is reviewed in \citep{Paesold2016conditions}. The present approach provides an alternative path to systematically analyse the full system of field equations.

\subsection{THMC acceleration waves}
We assume creeping flow in Eq. (\ref{eq:viscoplastic}), and there is therefore no effect of gravitational acceleration ($\rho \mathbf{F}=0$).  We will show that the meso-scale formalism identifies an alternate internal force density from within the considered material volume stemming from a local thermodynamic THMC force (e.g. $\nabla p$). This internal force integrates accelerations over the accelerations $\mathbf{a_M}$ of the micro-processes inside the continuum element and multiplying them by the average volume density. These accelerations stem from dissipative mechanisms (e.g. volume changes by phase transitions, fracture, etc.) inside the representative volume element \citep{Hu2019b}. For critical conditions they can cause acceleration waves propagating as creeping waves. Hadamard's jump conditions need to be extended for internal THMC variables $\mu$ such as temperature, porosity, permeability, viscosity etc..

Hadamard's jump conditions state that if  time derivatives ($\dot{\mathbf{F}}$, $\dot {\mathbf{v}}$, $\dot{\mu}$) and gradients $ \nabla \mathbf{F}, \nabla \mathbf{v}$, $\nabla \mu$) have jump discontinuities across the wavefront  then $\mathbf{F}, \mathbf{v}$ and $\mu$  are continuous functions of space. The compatibility condition relating jumps in rates of change of internal variables to jumps in gradients for all internal variables $\mu$ \citep{Perzyna1996} implies that the jump in the gradient of pressure inside the acceleration wave is constrained by

\begin{equation}
\label{eq:pdot}
[\nabla p] = -\frac{1}{c•} [\dot{p}]  .
\end{equation} 

Acceleration waves consider a step function (Eq. \ref{eq:A11}) where the stress-rate is discontinuous along the surface. The stress is, however, continuous across the wave front and the stress self-diffusion coefficient is also constant outside of the wave. Therefore, for modelling acceleration waves in a homogeneous material we can simplify Eq. (\ref{eq:A1}) further and assume constant bulk and shear viscosity outside the wave and assume continuity of stress across the acceleration wave. Noting that the traction in the direction of the normal vector $\mathbf{n}$  on the acceleration wave front is $\mathbf{F}= \mathbf{n} \cdot \boldsymbol{\sigma }$  it follows from Eq. (\ref{eq:A13}) that the jump in stress rate on the acceleration wave is \citep{Perzyna1996}:

\begin{equation}
\label{eq:stressrate}
\mathbf{n} \cdot [\boldsymbol{\dot \sigma }] = - c [\rho \mathbf{\dot v}] 
\end{equation}

Substituting the stress rate for the acceleration $\mathbf{\dot v}$  from Eq. (\ref{eq:stressrate}) and the pressure rate for the gradient of pressure from Eq. (\ref{eq:pdot}) and inserting the jump condition into Eq. (\ref{eq:viscoplastic}) it follows that

\begin{equation}
\label{eq:momentum3}
\frac {1}{c} \left[\frac{D(\mathbf{n} \cdot {\boldsymbol{\sigma})}} {Dt}\right]=-\frac{1}{c}[\dot{p} ]+[\frac{\dot{\boldsymbol{C}}}{c} \boldsymbol{\dot{\epsilon}}']-[\boldsymbol{C}\nabla (\boldsymbol{\dot{\epsilon}}')]+ [\frac{\dot{\zeta}}{c} \boldsymbol{\dot{\epsilon_0}}] -[\zeta \nabla\boldsymbol{\dot{\epsilon_0}}]  
\end{equation}  	

where $D(\cdot)/Dt$ denotes the material derivative with respect to the normal material velocity $\mathbf{v}$. If we define the magnitude of the wave speed in the normal reference system as  $w= \mathbf{w} \cdot \mathbf{n}$, then $c=w-\mathbf{v} \cdot \mathbf{n}$ is the local particle velocity of THMC accelerations in the acceleration wave relative to the normal material velocity.

Eq. (\ref{eq:momentum3}) allows us to draw some important conclusions for elasto-visco-plastic acceleration waves. (1) The first term on the right shows that the pressure rate divided by the wave velocity or the equivalent gradient of pressure plays an important role in acceleration waves. (2) The second and third term on the right imply that gradients of deviatoric strain-rates are related to rate changes of the stiffness tensor as implied by the jump condition of the internal variable inside the propagating wave.  Recall that the jump condition (Eq. \ref{eq:A11} or \ref{eq:pdot}) advects jumps in gradients of the internal variable around the propagating wavefront through a jump in the rate of change of the variable.  (3) The last two terms imply that the same is true for the volumetric strain rates and the rate of change of bulk viscosity. 


\section{ Multiscale cross-diffusion model}
So far we have only discussed the mechanical reaction-diffusion equation where the shear and bulk viscosities control the diffusion of stress.  For the multiphysics implementation it is convenient to think of diffusion of momentum and use the momentum diffusivity (kinematic viscosity) instead of the dynamic viscosity.  We therefore denominate $\zeta_M$  as the volumetric diffusion coefficient of pressure (kinematic viscosity). In the following we first formulate the reaction-diffusion equation in the classical way. That is to say that meso-scale cross-diffusion effects are neglected. We identify THMC-Turing patterns as multiscale energy eigenstates of the reaction-diffusion equations thus characterizing Prigogines dissipative structures, if they emerge.

In these formulations the viscous (M) mechanical pressure diffusion equation finds its counterparts in the equivalent thermal (T) Fourier- ,  (H) Darcy- and (C) Fick's- diffusion laws where the diffusion coefficients are indicated by the associated THMC subscript. The corresponding reaction rates are  ${{R}_{T}},{{R}_{H}},{{R}_{M}}$ and ${{R}_{C}}$, respectively.  It is common practice to ignore the meso-scale cross-diffusion kinetics introduced in the next section. We emphasize therefore the difference between large scale reaction rates $R_i$ and meso-scale reaction rates $r_i$ which consider the important effect of cross-diffusion. The two rates are identical in the infinite time scale limit as cross-diffusion can be eliminated adiabatically \citep{Biktashev2016}.  

In the adiabatic limit we obtain similar reaction-diffusion equations across a vast range of THMC diffusion length scales as tabulated in Table \ref{tab:THMC}. The reaction rates most often stem from different micro-processes at lower scale inside the considered continuum element which introduces cross-scale diffusion fluxes as shown in the next section. 

\begin{table}[h]
\centering
\caption{Generalized Thermodynamic Fluxes and Forces in a THMC coupled system (1-D)}
\label{tab:THMC}
\begin{tabular}{@{} l L L L @{} >{\kern\tabcolsep}l @{}}    \toprule
\textbf{\emph{Type \,\,\,\,}} & \textbf{\emph{Force\,\,\,\,\,\,\,\,\,\,\,\,\,\,\,\,\,}} &\textbf{\emph{Flux\,\,\,\,\,\,\,\,\,\,\,\,\,\,\,\,\,\,\,\,\,\,}}&\textbf{\emph{reaction-diffusion eq.}}&  \\\midrule
\rowcolor{black!5}[0pt][0pt] \textbf{T} & ${{F}_{T}}=\nabla T$ & ${{q}_{T}}=-\frac{DT}{Dt}$ & $\frac{DT}{Dt}={\zeta_T} \nabla^2 T+{{R}_{T}}$\\[0.07in] 
\rowcolor{black!10}[0pt][0pt] \textbf{H} &${{F}_{H}}=\nabla p_H$ &$ {{q}_{H}}=-\frac{D{{p}_{H}}}{Dt}$ & -$\frac{D{{p}_{H}}}{Dt}={{\zeta}_{H}} \nabla^2 p_H+ {\eta \boldsymbol{\dot{\epsilon}}'}-{{R}_{H}}$\\[0.07in] 
\rowcolor{black!20}[0pt][0pt] \textbf{M} &${{F}_{M}}=\nabla p_M$ & ${{q}_{M}}=-\frac{D{{{{p}}}_{M}}}{Dt}$ & -$\frac{D p_M}{Dt}=\zeta_M \nabla^2 {p_M}+\nabla (\boldsymbol{{C}}{{{\boldsymbol{{\dot{\epsilon}}'}}}})-{{R}_{M}}$\\[0.07in] 
\rowcolor{black!30}[0pt][0pt] \textbf{C}  & ${{F}_{C}}=\nabla C$ & ${{q}_{C}}=-\frac{DC}{Dt}$ & $\frac{DC}{Dt}=\zeta_C \nabla^2 C+{{R}_{C}}$\\[0.07in] 
\bottomrule
 \hline
\end{tabular}
\end{table}

In order to generalize the approach we propose that, in analogy to the quantum mechanical formulation for multiple particles, the multi-scale THMC wave operator $\hat{H}_{THMC}=\hat{H}_{kin}$  can be constructed through a linear superposition of Eq. (\ref{eq:Schroedinger1}) to obtain the quantum case for the interaction of an $i$-particle system travelling through free space as a sum of the energies characterized only by the kinetic Hamiltonian operator 
\begin{equation}
\hat{H}_{kin}=-\frac{\hbar^{2}}{2} \sum_{i=1}^{N} \frac{1}{m_{i}} \nabla_{i}^{2} , 
\label{Hamiltonian}
\end{equation}

where $\nabla_{i}^{2}=\frac{\partial^{2}}{\partial x_{i}^{2}}+\frac{\partial^{2}}{\partial y_{i}^{2}}+\frac{\partial^{2}}{\partial z_{i}^{2}}$.

The formulation can be extended below molecular length scales and one can consider electronic interaction with the associated Ohm's diffusion law, but we restrict the present discussion on the chemical length scale as the smallest entity. 


In the subsequent discussion we neglect the deviatoric terms in Eq. \ref{eq:momentum3} and retain only the scalar volumetric terms and reduce the equations to 1-D.  In order to introduce the meso-scale consideration we identify the reactive source terms ${R_T, R_H, R_M, R_C}$ as the local terms that in the case of acceleration waves turn a jump in the gradient in the variable into a rate of the variable (temperature, pressure, concentration). This term provides the convected pressure rate built up by internal accelerations. This relates to the local mass exchange processes according to Eq. \ref{eq:HillEinstein}.

\subsection{Cross-diffusion as a meso-scale formalism}
For evaluating the local mass exchange processes we use mixture theory \citep{Hu2019b}. We  consider two mass fractions $A$ and $B$ for mass exchange denoted by the $i^{th}$ and $j^{th}$  phase as an example. We identify $\dot \xi_i^{REV}$ as the large-scale Representative Elementary Volume (REV) for averaging of mass transfer rate from the phases $A$ to $B$ where $V_{REV}$, $V_A$, $V_B$ denote the REV volume and the volume of the  $i^{th}$ and  $j^{th}$  phase, respectively. $\dot \xi_i^{REV}$ defines the REV-scale averaging of mass exchange rate between the phases where the REV-scale source term of mass is obtained from the other species:

\begin{subequations}
\label{eq:local_global}
\begin{align}
\dot \xi_i^{REV} = - \frac{1}{V_{REV}}  \int_{V_{REV}} \dot \xi_j^{local},\\[0.2in]
\dot \xi_j^{REV} = - \frac{1}{V_{REV}}  \int_{V_{REV}} \dot \xi_i^{local},
\end{align}
\end{subequations}

where $\dot \xi_i^{local}$ and $\dot \xi_j^{local}$ denotes the mass exchange rate from the $A$ to $B$ phase and vice-versa. 

In the meso-scale formalism we need to consider information from the local scale processes in the THMC diffusion matrix and decompose the processes leading to the local mass production $\dot \xi_i^{local}$ and $\dot \xi_j^{local}$. 

In order to specify this further we define the global volume fraction of the  $A$-phase  as:
\begin{equation}
\label{eq:porosity}
\phi = \frac{V_A}{V_{REV}} = 1-\frac{V_B}{V_{REV}},
\end{equation}

Mass conservation at global scale for the phases $A$ and $B$ gives: 
\begin{subequations}
\label{eq:mass_balance}
\begin{align}
\frac{\partial [\rho_A V_A]}{\partial t} + \frac{\partial [\rho_A V_A  v_A]}{\partial x} = \dot \xi_A V_{REV}, \\[0.2in]
\frac{\partial [\rho_B V_B]}{\partial t} + \frac{\partial [\rho_B V_B v_B]}{\partial x} = \dot \xi_B V_{REV} .
\end{align}
\end{subequations}
$\rho_A$ and $\rho_B$ identify the density of the respective phases, while $v_A$ and $v_B$ their velocities in the direction of $x$ while $\dot \xi_A$ and $\dot \xi_B$ represent the volume averaged mass generation in the REV.

Substituting Eq.~\ref{eq:porosity} into Eq.~\ref{eq:mass_balance} and eliminating $V_{REV}$ gives
\begin{subequations}
\label{eq:mass_balance1}
\begin{align}
\frac{\partial [\rho_A \phi]}{\partial t} + \frac{\partial [\rho_A \phi v_A]}{\partial x}= \dot \xi_A^{REV}, \\[0.2in]
\frac{\partial [\rho_B (1-\phi)]}{\partial t} +\frac{\partial [\rho_B (1-\phi) v_s]}{\partial x} = \dot \xi_B^{REV}.
\end{align}
\end{subequations}

Repeating the same mass balance consideration for the local scale we obtain:
\begin{subequations}
\label{eq:mass_balance_local}
\begin{align}
 \dot \xi_B^{local} &= \frac{\partial [\rho_B (1-\phi^{local})]}{\partial t} + \frac{\partial [\rho_B (1-\phi^{local}) v_B]}{\partial x},  \\[0.2in]
\dot \xi_A^{local} &=  \frac{\partial [\rho_A \phi^{local}]}{\partial t} +\frac{\partial [\rho_A \phi^{local} v_A]}{\partial x},
\end{align}
\end{subequations}
%

Substituting Eq.~\ref{eq:mass_balance_local} into Eq.~\ref{eq:mass_balance1} via the upscaling law (Eq.~\ref{eq:local_global}), we have:

\begin{subequations}
\label{eq:mass_balance2}
\begin{align}
\frac{\partial [\rho_A \phi]}{\partial t} +\underbrace{\frac{\partial [\rho_A \phi v_A]}{\partial x}}_{Self-diffusion} + \underbrace{\frac{1}{V_{REV}}  \int_{V_{REV}}  \frac{\partial [\rho_B (1-\phi^{local}) v_B]}{\partial x}}_{Cross-diffusion} \nonumber \\
={- \frac{1}{V_{REV}}  \int_{V_{REV}} \frac{\partial [\rho_B (1-\phi^{local})]}{\partial t}}, \\[0.2in]
\frac{\partial [\rho_B (1-\phi)]}{\partial t} + \underbrace{\frac{\partial [\rho_B (1-\phi) v_B]}{\partial x}}_{Self-diffusion} + \underbrace{\frac{1}{V_{REV}}  \int_{V_{REV}}  \frac{\partial [\rho_A \phi^{local} v_A]}{\partial x}}_{Cross-diffusion} \nonumber \\
={- \frac{1}{V_{REV}}  \int_{V_{REV}} \frac{\partial [\rho_A \phi^{local}]}{\partial t}},
\end{align}
\end{subequations}

\subsubsection{Formulation of the THMC cross-diffusion matrix}
The concept of cross-diffusion is well known in chemistry. In a chemical system with just two species $A$ and $B$, for instance, cross-diffusion is the phenomenon, in which a flux of species $A$ is induced by a gradient of species $B$ \citep{Vanag}. In more general THMC terms, cross-diffusion is the phenomenon where a gradient of one generalised thermodynamic force drives another generalised thermodynamic flux.  Staying with the chemical example of species $A$ and $B$, we have in 1-D:

\begin{equation}
\label{eq:crossdiff1}
\begin{matrix}
 \frac{\partial {{C}_{A}}}{\partial t}={{\zeta}_{A}}\frac{{{\partial }^{2}}{{C}_{A}}}{\partial {{x}^{2}}}+{{L}_{AB}}\frac{{{\partial }^{2}}{{C}_{B}}}{\partial {{x}^{2}}}+{{r}_{B}}  \\ \\
 \frac{\partial {{C}_{B}}}{\partial t}={{\zeta}_{B}}\frac{{{\partial }^{2}}{{C}_{B}}}{\partial {{x}^{2}}}+{{L}_{BA}}\frac{{{\partial }^{2}}{{C}_{A}}}{\partial {{x}^{2}}}+{{r}_{A}} . \\ \\
\end{matrix}
\end{equation}

where $r_A$ and $r_B$ are the local source terms using the mesoscopic self-diffusion and cross-diffusion decomposition in Eq. (\ref{eq:mass_balance2}).
%
%

Following Eq. (\ref{Hamiltonian}) we can now generalize (\ref{eq:crossdiff1}) to include the full cascade of internal accelerations through multiscale coupling. Cross-diffusion allows coupling of accelerations from one classical REV-scale reaction-diffusion system, defined by the (self-)diffusive length scale  $\sqrt{\zeta_{T,H,M,C}t}$, to another. These (quantized) cross-diffusion coefficients link the gradient of a thermodynamic force ${C}_{j}$ of one THMC process to the flux of another kind. This allows the definition of a fully populated diffusion matrix as in:

\begin{equation}
\label{eq:crossdiffusionmatrix}
\frac{{D}{\mathbf{C}}}{{D}t}= 
\begin{bmatrix}  
   {\textcolor{red}{{\zeta}_{T}}} & {{L}_{TH}} & \textcolor{black}{{{L}_{TM}}} & {{L}_{TC}}  \\
   {{L}_{HT}} & \textcolor{red}{{{\zeta}_{H}}} & {{L}_{HM}} & {{L}_{HC}}  \\
   \textcolor{black}{{{L}_{MT}}} & {{L}_{MH}} & \textcolor{red}{{{\zeta}_{M}}} & {{L}_{MC}}  \\
   {{L}_{CT}} & {{L}_{CH}} & {{L}_{CM}} & \textcolor{red}{{{\zeta}_{C}}}  \\
\end{bmatrix} 
{{\nabla }^{2}}\mathbf{C}+{{r}_{i}} .
\end{equation}

The cross-diffusion coefficients introduce new cross-scale coupling length scales smaller than the self-diffusion scales  that provide the capacity to link different THMC processes. Ref. \citep{Hu2019b} shows an example of self-diffusion length scales in a geological application.

\subsubsection{Criterion for nucleation of Cross-Diffusion Waves}
An in-depth discussion of the criterion for instability and the type of instabilities for THMC systems will be the subject of a forthcoming contribution. Here, we summarize the basic method that is well established in the fields of mathematical biology and chemistry. A detailed discussion of the waveforms of cross-diffusion waves can also be found elsewhere \citep{Tsyganov2014}. 

The criterion for nucleation of cross-diffusion waves relies on assessing the dispersion relation of the eigenvalues of the characteristic matrix of a perturbed cross-diffusion-reaction equation \citep{Vanag}. The eigenvalues are functions of the square of a wavenumber of the perturbed state and identify the growth rate of the perturbations. This approach for deriving the mathematical criterion for nucleation of acceleration waves is hence evaluated from a small plane-wave  $\epsilon$-perturbation of Eq. (\ref{eq:crossdiffusionmatrix}) with 

\begin{equation}
\tilde{\mathbf{C}}(x, t)= \mathbf{C}_{0} (1+\epsilon) \mathrm{e}^{\lambda t+i\left(k x\right)}
\label{instability}
\end{equation}

The characteristic matrix of the thus perturbed Eq. (\ref{eq:crossdiffusionmatrix}) allows assessing the stability of the system. Accordingly, all eigenvalues of the charcteristic matrix must be real and positive, and hence the determinant of the matrix  must be larger than zero.  For determinants smaller than zero cross-diffusion waves are expected to propagate as quasi-solitons \citep{Tsyganov2014}.  A worked example for hydromechanical cross-diffusion waves can be found in reference \citep{Hu2019b} and an application to the nucleation of earthquake instabilities is discussed in ref. \citep{Regenauer2019}.
 
 \section{Conclusions}
We have presented a simple mesoscale theory for the multi-physics of material instabilities and shown its link to classical theories of localisation phenomena in continuum mechanics. The approach reveals that instabilities based on the volumetric response of the material are fundamentally important but have been overlooked in earlier theories (except for ref. \citep{Veveakis2015cnoidal,Hu2019b}).  We have also shown that incompatibilities of local accelerations with the overall stress field lead to the nucleation of cross-diffusion waves which travel in a particle-like state with characteristic material velocities $c$ defined by the competition of local reaction-diffusion processes at the propagating wavefront. These velocities characterise the progress of internal material time-scales for the formation of multiscale space/time dissipative structures. These internal material clocks are here introduced as a  multiscale THMC cascade for coupling the physics of the very small to the very large. 

Cross-diffusion waves have first been discovered in Laser Optics where they show anomalous dispersion patterns that, unlike solitons, come in discrete portions \citep{Paschotta}. They have consequently been introduced as quasi-solitons. Since the speed and shape of quasi-solitons are fixed, the reflections and collisions, when they happen, are always 'quasi-elastic' {Biktashev2016,DissipativeAccelerationWave}.  In spite of their nucleation through discrete internal micro-dissipative mechanisms, cross-diffusional waves also show proper wave-like behaviour and can penetrate through each other and reflect from boundaries \citep{Regenauer2019}. However, unlike true solitons their amplitude and speed is not controlled by initial conditions but by material properties. 

The effect of cross-diffusion  is to trigger cross-diffusion waves for critical conditions. They form by THMC feedback as discrete material instabilities which  can be either observed as locally discrete failure or as  damage waves whose modulus of probability amplitude squared provides the probability density distribution of local dissipative structures quantifying uncertainty for the location of failure. We have shown that they can be decomposed into cross-diffusional $S$- and $P$-waves and have discussed a THMC multiphysics implementation, where cross-diffusion waves appear as quasi-soliton waves for critical conditions identified from a perturbed Eq. (\ref{instability}). Their role in coupling length-scales through multiphysics is not restricted to THMC applications. As a generic theory the cross-diffusion approach has been shown to be mathematically identical to the Schr\"odinger equation for the motion of a quantum state particle and it bears the same uncertainty relationship. 

\section{Acknowledgments} 
This work was supported by the Australian Research Council (ARC DP170104550,  DP170104557) and the strategic SPF01 fund of UNSW, Sydney. We would also like to acknowledge constructive feedback of Mike Trefry.
%

\newpage

\section{References}
\bibliography{Uncertainty}

\end{document}